\documentclass[aps,prb, twocolumn,groupedaddress]{revtex4-1}
\newcommand{\up}{\uparrow}
\newcommand{\dw}{\downarrow}

\begin{document}


\title{Spin effects induced by thermal perturbation \\
 in a normal metal/magnetic insulator system}


\author{ I.I. Lyapilin, M.S. Okorokov, V.V. Ustinov}
\email {Okorokovmike@gmail.com}
\affiliation{Institute of Metal Physics, UD RAS, Ekaterinburg, 620137, Russia}


\date{\today}

\begin{abstract}
Using one of the methods of quantum  nonequilibrium statistical physics  we have investigated the
spin transport transverse to the normal metal/ferromagnetic insulator interface in hybrid
nanostructures.  An approximation of the effective parameters, when each of the interacting
subsystems (electron spin, magnon, and phonon) is characterized by its own effective temperature
have been considered. The generalized Bloch equations which describe the spin-wave current
propagation in the dielectric have been derived.  Finally, two sides of the spin transport "coin" have been revealed: the diffusive nature
of the magnon motion and  magnon relaxation processes,  responsible for the spin pumping and the spin-torque effect.

\end{abstract}

\maketitle


\section{Introduction}

One of the central issues of spintronics is the generation and control of spin currents in solids. There are different methods to implement the spin current: optical, magnetic, and using an electric current. The latter is especially important for use in devices \cite{Wol} when spin-polarized charge carriers are injected from a ferromagnetic material to a non-magnetic. At that, in the nonmagnetic material at the spin diffusion length, spin accumulation arises. With an external
perturbation usually affecting the kinetic degrees of freedom, the interaction between translational (kinetic) and spin degrees of freedom plays the main role in the formation of the spin response to the external perturbation. Combined electric dipole resonance can serve as an example of such an effect. In this case, the interaction of conduction electrons and an alternating electric field results in resonance at the Zeeman frequency \cite{UFN}. Another example of such a
response is the spin Hall effect (SHE) \cite{Djak,Hirsh} that is exhibited as a spin current perpendicular to both the normal current and the spin accumulation. There also exist mechanisms of interaction with external fields whose energy is simultaneously transferred to both electronic subsystems (kinetic and spin) \cite{Lyap}. It has been found that thermal perturbations may also cause the spin effects to occur. The first effect that has opened a new direction in spintronics - the influence of thermal perturbations on spin effects, is the spin Seebeck effect (SSE)  \cite{Ucida,Jawor}. It has been turned out to be inherent in conducting crystals of $Ni_{81}Fe_{19}$. Afterwards, the SSE could be observed in various materials such as $(Ga,Mn)As $\,\cite{Jawor} semiconductors  and metallic $Co_2MnSi$ ferromagnets  \cite{Uci-2}. Besides, later the spin Nerst effect (or the thermal spin Hall effect), the spin Peltier effect, and others have been discovered \cite{Ma,Liu,Che,Dur,Grav,Hu}. As noted in the work \cite{Maek}, there is much common between the spin effects implemented in an electric or an inhomogeneous temperature fields. So, in spintronics, studying the interaction
between charge and spin currents, the new direction appears - spin caloritronics \cite{Ed,Bau}. As far back as in the late 20th century, a few theoretical aspects of these interactions were discussed \cite{Jon}.

Studying of the SSE in a non-conducting magnet in the system of a nonmagnetic conductor / magnetic
insulator (NM/FI) $La\,Y_2\,Fe_5\,O_{12}$  \cite{Uci-3} has shown that this effect cannot be
described by standard approaches as regards a description of thermoelectric effects \cite{Jon}. As
distinct from conducting crystals where the transfer of the spin angular momentum is due to band
charge carriers, the spin Seebeck effect can be realized in non-conducting magnetic materials
through the excitation of a localized spin system. In this case, the excitations (magnons)
underlying the spin-wave current  causes the transfer of the angular momentum. Thus, conducting
crystals and a nonconductive magnet differ from each other in type of the spin current, but namely
- spin-wave one. It is a new type of the spin current. The system of localized spin  can be
deviated from its equilibrium state in various ways. In the experiment  \cite{Uci-3}, this has been
achieved by passing an electric current through the nonmagnetic conductor  $Pt$, which results in
the spin accumulation and the spin current in it. The interaction of the spin-polarized electrons
with the localized spins at the interface (NM/FI) is accompanied by the creation (or annihilation)
of magnons, which in turn gives rise to a perturbation of the magnetic subsystem. Due to small
spin-wave damping, the spin-wave current propagates at much greater distances than the electron
spin current. This fact makes the effect possible to apply in practice \cite{Tak,Bak}.

An important role in the study of thermal perturbations is played by a lattice (phonons). Indeed, a
inhomogeneous temperature field may cause a deviation of both the localized spins subsystem $(m)$
and the phonon subsystem $(p)$ from their equilibrium state. If the non-equilibrium state of each
subsystem is characterized by an effective temperature $T_i,\,\,i=m,p$, the non-equilibrium state
of the system as a whole can be clearly described by a set of parameters $T_i$. The difference in
the effective temperatures of the subsystems can lead to the realization of the effect - the
transfer of the  angular momentum  from the magnetic subsystem to the lattice or to the electron
spin subsystem \cite{Nak,Xiao} and vice versa. Thermal perturbations are also responsible
for the drag effects, whose role is essential in a range of not very high temperatures
\cite{Gur,Blat}. As the authors \cite{Uci-4} believe, it is the thermal perturbations
which produce the observable anomaly in the temperature dependence of the SSE.

The spin effects in non-conductive magnetic materials under thermal perturbations were studied in
several papers \cite{Ada-1,Ada-2,Xia,Nak}. In this case, the theoretical
description boiled largely down to the consideration of the evolution of the localized moment
subsystem. Given thermal fluctuations derived from the fluctuation-dissipative theorem, the authors
of the works mentioned above modeled the localized moments' dynamics by the phenomenological
Landau-Lifshitz-Gilbert equation. As to the spin density dynamics in a nonmagnetic material, it is
described by the Bloch equation with phenomenological spin relaxation frequencies. A spin current
that was injected into the non-magnetic material $(NM)$, was calculated as the average value of the
rate of change of the spin density in a nonmagnetic material. Spin-charge kinetics in the systems
$(F/N)$, $(F/N/F)$ under a temperature gradient was thoroughly covered in the paper  \cite{Sha} in
terms of a simple phenomenological model and the simplest approximations.

The study of the spin-thermal effects requires a correct description of the thermal perturbation.
There are several possible scenarios accounting for such perturbations. For example, the reaction
of weakly nonequilibrium systems on a disturbance of the thermal type can be represented as Fourier
transform of time correlation functions of appropriate flux operators over the statistical
equilibrium state of the system  \cite{Ku}. The structure of admittances of such a type is
analogous to expressions for the transport coefficients arising in the theory of equilibrium
systems as the response to a mechanical perturbation. These expressions can be represented as an
additional summand in the system's Hamiltonian. That is why the response to a thermal perturbation
is often found using indirect methods. Such cases require introducing fictitious external forces
identical to thermal perturbations exerted on the system \cite{Lut}. Response to the thermal
perturbations can also be calculated based on the Onsager hypothesis \cite{On} about of the nature
of fluctuation damping or by using the local-equilibrium distribution as an initial condition for
the statistical operator of the system to be found \cite{Gr}.

The response of weakly nonequilibrium systems to the thermal perturbations can be universally
constructed through the method of the nonequilibrium statistical operator (NSO) and its
modifications \cite{DNZ}. The latter represent the NSO as a functional of the local-equilibrium
distribution. The method allows calculating the response of not only equilibrium but also strongly
nonequilibrium systems described in terms of roughened macroscopic variables. Within this method,
the kinetic coefficients are expressed via the Fourier transform of time correlation functions over
the statistical distribution which describes the unperturbed nonequilibrium process. Naturally, in
this case, electron scattering needs to be taken into account.

In the given paper, we consider the spin-thermal effects in a normal conductor/magnetic insulator,
using the NSO method to describe thermal perturbations. The article is organized as follows. The
first part formulates a model at hand, represents a Hamiltonian of the system and introduces basic
operators and their microscopic equations of motion. The second part of the work covers
constructing both the nonequilibrium entropy operator including the thermal perturbations of the
system and the NSO. The third section of the article focuses on an analysis of macroscopic
equations describing the spin-thermal effects.

\section{The  Hamiltonian}

Our model consists of a normal conductor $(NM)$ and a ferromagnetic insulator $(FI)$. We hold that,
in the first conductor, the spin accumulation takes place. It can be obtained and implemented in
various ways, for example by using the spin Hall effect. The conduction electrons of the normal
metal are coupled with the localized spin subsystem of the ferromagnetic insulator through exchange
interaction. The inelastic scattering of the electrons by the localized spins, accompanied by
magnon emission or absorption, alters the electron spin orientation and unbalances the localized
spin system.  Consider the magnon scattering mechanism by the example of the magnon-phonon
interaction. We assume the ferromagnetic insulator to be in a non-uniform temperature field. Let
the system of conduction electrons in the normal conductor consist of two subsystems: the kinetic
and spin one. In this case, the first is characterized by an equilibrium temperature $T$, and the
second (the spin subsystem) - by a temperature $T_s$. The symbol $T_m$ designates a temperature of
the localized spin subsystem, $T_p$ - a temperature of the lattice (phonons).

The Hamiltonian of the system $(NM/FI)$  can be represented as $H=H_{NM} + H_{FI} +H_L$. Here
\begin{eqnarray}\label{1}
H_{NM} =  \int d{\bf x}\, (\,H_k({\bf x})+  H_s({\bf x})\,),\qquad\qquad\qquad\nonumber\\
H_k({\bf x}) = \sum_j \{\frac{p_j^2}{2m},\delta({\bf x}-{\bf x}_j)\},\nonumber\\
H_s({\bf x}) = -\hbar\omega_s\,\sum_j\,s^z_j\,\delta({\bf x}- {\bf x}_j).
\end{eqnarray}
The integration is over the volume occupied by $(NM). $ $s^z_j$   and $p^\gamma_j$ are components
of the spin  and angular momentum of $j$-th electron operators, respectively.
$\omega_s=g_s\mu_0H/\hbar$ is the Zeeman precession frequency of free electrons in an external
magnetic field directed along the axis $z$. ($g_s,\,\mu_0$ are the effective spectroscopic
splitting factor for electrons and the Bohr magneton, respectively); $\{A, B\} = (AB + BA) / 2$.
\begin{equation}\label{2}
H_{FI} =  \int d{\bf x}\, (H_m({\bf x}) +H_{ms}({\bf x}) )
\end{equation}
is the Hamiltonian of of the localized spins subsystem .  $H_m({\bf x})$   is the energy density
operator of the magnetic subsystem. It involves a sum of the exchange (over the nearest neighbors)
$H_{SS}$ and Zeeman energy $H_S$.

\begin{equation}\label{4}
H_{SS}= -J\sum_{j\delta}\,S_j S_{j+\delta},\qquad H_S = -\hbar\omega_m\sum_j\,S_j^z
\end{equation}
$J$ is the exchange integral. $\omega_m = g_m\mu_0 H/\hbar.$  $H_{ms}({\bf x})$  is the energy
density operator of interaction between the $s$ and $m$  subsystems at the interface. Then

\begin{equation}\label{7}
 H_{ms} = -J_0\,\sum_j\int\,d{\bf x}\, {\bf s}({\bf x})\,{\bf S}({\bf R}_j)\,\delta({\bf x}-{\bf R}_j),
 \end{equation}
where $J_0$  is the exchange integral,  $S({\bf R}_j)$  being the operator of the localized spin
with the coordinate ${\bf R}_j$  at the interface. The integration in (\ref{2}) is over the volume
occupied $(FI)$.

$H_L$ is the lattice hamiltonian
 $$H_L = \int d{\bf x} \,(H_p({\bf x}) + H_{pm}({\bf x})\,)$$
$H_p({\bf x})$  is the energy density operator of the phonon subsystem. $H_{pm}({\bf x})$ is the
energy density operator of interaction between the localized spins and phonons. In the future, we
are interested in the evolution of the magnetic subsystem, so the scattering of conduction
electrons by phonons for simplicity omitted from consideration.

\section*{\bf  The entropy operator}

To analyze the kinetics of the spin-thermal effects, we employ a scheme developed in the NSO method
applied to the case of a small deviation of the system from the equilibrium Gibbs distribution
$\rho_0=\exp\{-S_0\}$. The entropy  $S_0$ of the equilibrium system with the Hamiltonian $H$ can be
written as
$$S_0= \Phi_0 +\beta (H_k+H_s - \mu\, n) + \beta (H_m + H_p+H_{ms}+  H_{mp}),$$
where  $\beta^{-1} = T$  is the equilibrium temperature of the system.

In terms of average densities, to the non-equilibrium state of the system there corresponds the
entropy operator
\begin{eqnarray}\label{8}
S(t)= \Phi(t) + \int d{\bf x}\,\{\beta (H_k({\bf x}) - \mu^\alpha({\bf x},t) n^\alpha({\bf x},t))+
 \nonumber\\
+ \beta_s({\bf x},t)H_s({\bf x},t) + \beta\,H_p({\bf x},t)+\nonumber\\+\beta_m({\bf x},t) (H_m({\bf
x},t)+H_{ms}({\bf x},t) +H_{pm}({\bf x},t))=\nonumber\\= S_0 + \delta S(t).\quad\quad
\end{eqnarray}
$\Phi(t)$  is the Massieu-Planck functional. $\beta_s({\bf x},t)$, $ \beta_m({\bf x},t) $  are the
local-equilibrium values of the inverse temperature of the $(s),\,(m)$ subsystems, respectively.
$\mu^\alpha({\bf x},t)$ is the local-equilibrium value of the chemical potential of electrons with
spin $\alpha =\up,\dw$.
$$n^\alpha({\bf x}) = \sum_j\delta ({\bf x} - {\bf x}^\alpha_j)$$
is the  operator of the particle number density  with the  spin $\alpha$.\, $n({\bf x}) =
n^\up({\bf x}) + n^\dw({\bf x})$.

The operator $\delta S(t)=\int d{\bf x} \delta S({\bf x},t)$ describes the system deviation from
its equilibrium state.
\begin{eqnarray}\label{9}
\delta S(t)=\Delta\int d{\bf x} \,\{\delta\beta_s(x,t)H_s(x) - \beta\,\delta\mu^\alpha({\bf
x},t)\,n^\alpha({\bf x}) +\nonumber\\\qquad +\delta\beta_m({\bf x},t) (H_m({\bf x})+H_{ms}({\bf
x})+H_{pm}({\bf x}))\},\qquad
\end{eqnarray}
$$ \Delta A = A - <A>_0,\quad <\ldots>_0 = Sp\,(\ldots\rho_0).$$
$$<\ldots>^t = Sp\,(\ldots \rho(t)),\quad \rho(t) =\exp\{-S(t)\}.$$
$$\delta\mu^\alpha(x,t)= \mu^\alpha(x,t) -\mu,\quad \delta\beta_i(x,t)= \beta_i(x,t)-\beta.$$
$(i = s,\,m).$ The quantities  $ \delta\mu^\alpha({\bf x},t),\delta\beta_m({\bf x},t)$ have the
meaning of non-equilibrium additions to the chemical potential   $\mu$ and deviations of the
magnetic subsystem temperature from the equilibrium temperature $\beta^{-1}=T$.

In the linear approximation the deviation from the equilibrium, the NSO (or the density matrix)
$\rho(t)$   can be written as follows \cite{DNZ}
\begin{eqnarray}\label{10}
\rho(t) =\rho_q(t) + \int\limits_{-\infty}^0\,dt_1\,e^{\epsilon t_1}\int\limits_0^1d\tau\,
\rho_0^\tau\dot{S}(t+t_1,t_1)\rho_0^{-\tau}\,\rho_0.\qquad
\end{eqnarray}
Here $\rho_q(t)=\exp\{-S(t)\}$  is the quasi-equilibrium statistical operator, $\dot{S}(t)$ being
the entropy production operator:
$$\dot{S}(t)=\delta\dot{S}(t) = \frac{\partial S(t)}{\partial t} +\frac{1}{i\hbar}[S(t), H].$$
A further algorithm for constructing the operator $\rho(t)$  reduces to finding the entropy
production operator. Commuting the operators $n^\alpha({\bf x}),\, H_s({\bf x})$   with the
Hamiltonian (\ref{1}), we find the operator equations of motion
\begin{equation}\label{11}
\dot{n}^\alpha({\bf x})=- {\bf \nabla}\,I_{n^\alpha}({\bf x})+ \dot{n}^\alpha_{(ms)}({\bf x}),
 \end{equation}
 \begin{equation}\label{12}
\dot{H}_s({\bf x}) = - {\bf \nabla}\,I_{H_s}({\bf x}) + \dot{H}_{s(ms)}({\bf x}).
 \end{equation}
Here $I_{n^\alpha}({\bf x})= \frac{1}{m}\sum_j\{p_j,\delta({\bf x} - {\bf x}^\alpha_j)\}$   is the
flux density of particles with spin  $\alpha$.   $\dot{n}^\alpha_{(ms)}({\bf x})$ determines the
rate of change of the number of electrons with  spin $\alpha$ due to the spin-flip electron
scattering by the $(m)$ -subsystem.

Similarly, $ I_{H_s}({\bf x})=-\hbar\omega_s \frac{1}{m}\sum_js^z_j\{p_j,\delta( {\bf x}- {\bf
x}^\alpha_j)\}$  determines the flux density of the Zeeman energy.   $\dot{H}_{s(sm)}({\bf x})$ is
the rate of change  of the electrons  $s$-subsystem  local energy due to their interaction with
magnons $(m)$.
\begin{equation}\label{13}
\dot{A}_{\lambda(\gamma)}({\bf x})= (i\hbar)^{-1}[A_\lambda({\bf x})\,, H_\gamma].
\end{equation}
Revealing the spin index explicitly, we write the expression for the spin current density $J_s({\bf
x})$:
\begin{eqnarray}\label{14}
J_s({\bf x})=(1/2)\{\dot{n}^\up({\bf x})-\dot{n}^\dw({\bf x})\}=-\nabla\,I_{s^z}({\bf x})+
\dot{s}^z_{(ms)}({\bf x}),
\qquad\nonumber\\
I_{s^z}({\bf x}) =( I^\up_s({\bf x}) - I^\dw_s({\bf
x}))/2,\qquad\qquad\nonumber\\\dot{s}^z_{(ms)}({\bf x}) = (\dot{n}^\up_{(ms)}({\bf x}) -
\dot{n}^\dw_{(ms)}({\bf x}))/2.\qquad\qquad
 \end{eqnarray}

From the expression (\ref{14})  it follows that $J_s({\bf x})$ contains two parts: collisionless
$I^z_s({\bf x})$,  and collisional $\dot{s}^z_{(ms)}({\bf x})$.  The former is due to the flux of
particles with different spin orientations, the latter is determined by the spin-flip scattering.

Let us turn to the consideration of the magnetic subsystem. Using the method of Holstein-Primakov
\cite{Mat}, the hamiltonian  of localized spins subsystem can be rewritten with spin-wave (magnon)
variables (using to the operators creation $b^+_{\bf k}$ and annihilation $b_{\bf k}$). Expressed
in terms of the magnon operators, the Zeeman and exchange interactions are of the form
\begin{equation}\label{5}
H_S = -\hbar\omega_dNS + \hbar\omega_d \sum_k\,b^+_{\bf k}\,b_{\bf k}.
 \end{equation}
\begin{eqnarray}\label{6}
H_{SS}\simeq -J N z S^2 +\sum_k\,2JzS(1-\gamma_{\bf k})b^+_{\bf k}\,b_{\bf k},\nonumber\\
\gamma_{\bf k}=(1/z)\sum_\delta\exp\{i{\bf k}\,{\bf\delta}\}.
 \end{eqnarray}
Here $N$ is the number of localized moments with the spin $S, z$ is the number of the nearest
neighbors.  $b^+_{\bf k}\,b_{\bf k}$ are the creation and annihilation operators for magnons with
the wave vector ${\bf k}$. For $|{\bf k}\,{\bf  \delta}|<<1$ , we have $z(1-\gamma_{\bf k})\approx
(1/2)\sum_\delta ({\bf k}\,{\bf  \delta})^2$ , i.e. $\omega_{\bf k} = JS\sum_\delta (k\delta)^2$.
For body-centered and face-centered cubic lattices with a lattice constant equal to $a$, we have
$\omega_{\bf k} = 2J S (k a)^2$.  In other words, the contribution of the exchange interaction for
the magnon frequency has the same form as the de Broglie dispersion law for free particles of mass
$m^*$
  $$ \omega_k = \frac{\hbar}{2m^*}\,k^2,\mbox{где}\,\, m^* = \frac{1}{2JSa^2}.$$
Thus, regarding the magnon gas as free, we arrive at
 $$H_m = \sum_{\bf k} \varepsilon(k) b^+_{\bf k}\,b_{\bf k},\,\, \,\mbox{где}\,\,\, \varepsilon(k) =
 \frac{\hbar^2 k^2}{2 m^*}.$$
This expression can be interpreted as the sum of the energies of the quasiparticles-ferromagnons
having the quasimomentum $ {\bf P}$ with their own effective mass $m^*$ and the magnetic moment
\cite{Tur}.

The exchange interaction (\ref{7}) includes both the elastic electron scattering by the localized
spins and the inelastic scattering process, with the former preserving the spin orientation and the
latter happening with the electron spin-flip and the creation or annihilation for magnons. Using
the creation and annihilation operators for electrons and magnons, we rewrite the inelastic part of
the exchange interaction in the form

\begin{equation}\label{7a}
H_{ms} = -J^*\,\sum_{{\bf k},{\bf k}',{\bf q}}\,\{b^+_{\bf q}\,a^+_{{\bf k}\up}\,a_{{\bf k}'\dw} +
b_{\bf q}\,a^+_{{\bf k}'\dw}\,a_{{\bf k}\up} \}\, \delta_{{\bf k}',{\bf k}+{\bf q}},
 \end{equation}
where  $a^+_{{\bf k}\alpha}\,(a_{{\bf k}\alpha})$   are the creation (annihilation) operators for
electrons with a certain spin value $\alpha=\up,\,\dw$.   The inelastic part of the exchange
interaction ($H_{ms}$) leads to the angular momentum transfer between electrons of the normal metal
and magnons of the ferromagnetic insulator. Obviously, non-equilibrium of one of the subsystems
(electronic or magnon) makes the magnon/spin current via the interface. At the same time, the total
angular momentum conservation law provides the boundary condition for the current at the
metal/insulator interface: the continuity of the spin current.

Thus, the equation of motion for the magnetic subsystem can be written as
\begin{equation}\label{15}
\dot{H}_m({\bf x})= -\nabla\,I_{H_m}({\bf x})+\dot{H}_{m(ms)}({\bf x}) + \dot{H}_{m(mp)}({\bf x}).
 \end{equation}
Here   $ I_{H_m}({\bf x})=-\hbar\omega_m\, I_{S^z}({\bf x})$  is the flux density of the magnon
energy in the $(m)$ -subsystem. The remaining summands on the right-hand side of the equation are
responsible for the magnon-phonon scattering processes.

Finally, the equation of motion for the lattice subsystem has the form:
\begin{equation}\label{17}
\dot{H}_p({\bf x}) = -\nabla\,I_{H_p}({\bf x})+\dot{H}_{p(mp)}({\bf x}) .
 \end{equation}
Substituting the above equations of motion into the entropy production operator $\dot{S}(t)$ and
integrate by parts the summands containing the divergence of fluxes. Having discarded the surface
integrals, we write the entropy production operator in the form
\begin{eqnarray}\label{18}
\dot{S}(t)=\Delta\int d{\bf x}\{-\beta\,I_{n^\alpha}({\bf x})\nabla\mu^{\alpha}({\bf x},t)-
\beta\delta\mu^\alpha({\bf x},t)\dot{n}^\alpha_{(ms)}({\bf x})+\nonumber\\
+I_{H_m}({\bf x})\nabla\beta_m({\bf x},t) + (\beta_s({\bf x},t)-\beta_m({\bf
x},t))\dot{H}_{s(ms)}({\bf x})+\nonumber\\
+ (\beta_m({\bf x},t)-\beta)\dot{H}_{m(pm)}({\bf x})\}.\qquad\qquad
\end{eqnarray}
Revealing the spin index in the first two summands on the right-hand side of the formula
(\ref{18}), we get
\begin{eqnarray}\label{19}
I_{n^\alpha}({\bf x})\nabla\mu^{\alpha}({\bf x},t)= I_{s^z}({\bf x})\,\nabla\,\mu_s({\bf x},t).
\end{eqnarray}
\begin{eqnarray}\label{20}
\delta\mu^\alpha({\bf x},t)\dot{n}^\alpha_{s(ms)}({\bf
x})=\qquad\qquad\nonumber\\
=\mu_s({\bf x},t)\dot{s}^z_{s(ms)}({\bf x})= -\mu_s({\bf x},t)\dot{S}^z_{m(ms)}({\bf x}). \quad
    \end{eqnarray}
Here  $\mu_s = \mu^\up - \mu^\dw.$ It is obvious that the expression  (\ref{20}) provides the
boundary conditions for the collisional part of the spin current at the interface. The expressions
(\ref{19}), (\ref{14})  imply that the heterogeneity in the spin accumulation distribution governs
the collisionless part of the spin current.

Before proceeding to constructing the equations for the averages, we show that the entropy
production operator  $\dot{S}(t)$ can be represented both through the spin accumulation $\mu_s({\bf
x},t)$ and through the spin temperature $\beta_s({\bf x},t)$.  For this purpose, we find a linear
relationship between the deviations of thermodynamic coordinates and thermodynamic forces from
their equilibrium values. In this approximation, it is sufficient to put $\rho(t)\sim\rho_q(t)$ and
expand the quasi-equilibrium operator $\rho_q(t)=\exp\{-S(t)\}$  in powers of $\delta S(t)$ . Then
we come up with the result
\begin{eqnarray}\label{23}
\delta<H_s({\bf x})>^t =-\int d{\bf x}'\{\delta\beta_s({\bf x}',t)(H_s({\bf x});H_s({\bf x}'))_0-
\nonumber\\-
\beta\delta\mu^\alpha({\bf x}',t)(H_s({\bf x});n^\alpha({\bf x}'))_0\},\qquad\qquad\nonumber\\
\delta<n({\bf x})>^t =-\int d{\bf x}'\{\delta\beta_s({\bf x}',t)(n({\bf x});H_s({\bf x}'))_0
-\nonumber\\
-\beta\delta\mu^\alpha({\bf x}',t)(n({\bf x});n^\alpha({\bf x}'))_0\},\qquad\qquad
\end{eqnarray}
where $$\delta < A >^t = < A >^t - < A >_0,$$ $$ (A,B)_0=\int\limits_0^1 d\lambda
Sp\{A\rho^\lambda_0\,\Delta B\,\rho^{1-\lambda}_0\}.$$ Going over to the Fourier components of the
spatial coordinates and taking into account that $< n >^t  = < n >_0$, we obtain
\begin{eqnarray}\label{24}
\mu_s({\bf q},t) =  -  \frac{\hbar\omega_s\,}{\beta} \,\frac{(n({\bf q}),s^z(-{\bf q}))_0}{( n({\bf
q}),n(-{\bf q}))_0}\,\delta\beta_s({\bf q},t).
\end{eqnarray}
Equation (\ref{24}) defines the relationship between the spin temperature $T_s$ and the spin
accumulation $\mu_s$. In the long-wavelength limit ${\bf q}=0$ and the steady state, we obtain
\begin{eqnarray}\label{24a}
\mu_s =  -  \hbar\omega_s\,(1-T/T_S).
\end{eqnarray}
Substituting the expression (\ref{24}) into the equation for $\delta<~H_s({\bf x})>$ , we get
\begin{eqnarray}\label{25}
\delta< s^z({\bf q}) > = \delta\beta_s({\bf q},t)\,\hbar\omega_s\,C_{zz}({\bf q}),\nonumber\\
C_{zz}({\bf q}) = (s^z({\bf q}),s^z(-{\bf q}))_0-\frac{(n({\bf q}),s^z(-{\bf q}))_0^2}{( n({\bf
q}),n(-{\bf q}))_0}.
\end{eqnarray}
The desired relationship between the correlation function $C_{zz}$   and the differential
paramagnetic susceptibility $\chi_s$ \cite{Bik}  of the electron gas is given by
\begin{eqnarray}\label{26}
\chi_s = \frac{\partial}{\partial H}\,g_s\mu_b<s^z>= \beta(g_s\mu_b)^2\,C_{zz}.
\end{eqnarray}

\section{\bf Macroscopic equations}

Let us construct the equations for the averages, which have the meaning of local conservation laws of the average energy density for the $(s)$ and $(m)$ subsystems and of the particle number density. Inserting the entropy production operator  into the expression for the NSO (\ref{10}), we
average the operator equation (\ref{14})  for the spin current  $<J_s({\bf x})> = (d/dt)<s^z({\bf x})>$. Then we arrive at
\begin{eqnarray}\label{27}
 <\overline{J_s({\bf x})}>=\int d{\bf
x}'\,\{D^{\gamma}_{s^z\,s^z}({\bf x},{\bf x}')\beta\,{\bf \nabla}^\gamma\mu_s({\bf x}' +\nonumber\\
+(\beta_m({\bf x}')-\beta_s({\bf x}'))\,\hbar\omega_s\,L^{z}_{z,(ms)}({\bf x},{\bf x}' )+\nonumber\\
+\,\beta\,\mu_s({\bf x}',t+t_1)\,L^{z}_{z,(ms)}({\bf x},{\bf x}' )\},\qquad\quad\nonumber\\
 \end{eqnarray}
 where
\begin{eqnarray}\label{27a}
L^z_{z,(ms)}({\bf x},{\bf x}')  = \int\limits_{-\infty}^0dt'\,e^{\epsilon t'} \,(\dot{s}^z_{(ms)}({\bf x})\,;\,\dot{s}^z_{(ms)}({\bf x}',t')),
 \nonumber\\
D^{\gamma}_{s^z\,s^z}({\bf x},{\bf x}')=\int\limits_{-\infty}^0dt'\,e^{\epsilon
t'}\,\nabla^\lambda\,(I^\lambda_{s^z}({\bf x}); I^\gamma_{s^z}({\bf x}',t'))\,.\qquad
 \end{eqnarray}
Here  $\overline{A}$ denotes time averaging, $\overline{\beta_i({\bf x},t+t')} = \beta_i({\bf x}), (i=s,m);$ \,$\overline{\mu({\bf x},t+t')}=\mu({\bf x}). $

We have derived the generalized Bloch equation that describes the motion of the spin magnetization density - the spin diffusion and relaxation processes.  The first summand on the right-hand side of the equation reflects the former process; the second summand is responsible for the latter that
caused by the electron-magnetic impurity interaction at the interface. In doing so, we have made an allowance for the effects of temporal and spatial dispersion of the spin diffusion tensor and spin relaxation frequency.

Performing the Fourier transform over the coordinates and time and taking into account formulas (\ref{24}), (\ref{25}), we represent the right-hand side of expression as
\begin{eqnarray}\label{28}
-\,\{q^\gamma\,q^\lambda\, D_{zz}^{\gamma\lambda} ({\bf q},\omega) +\nu_{zz} ({\bf
q},\omega)\}P({\bf q})\,\delta<s^z ({\bf q},\omega)>,\nonumber\\
  P({\bf q}) =
\frac{(n({\bf q}),\,s^z(-{\bf q}))_0}{( n({\bf q}),n(-{\bf q}))_0}.\qquad\qquad
\end{eqnarray}
Here
\begin{equation}\label{29}
\nonumber\\D_{zz}^{\gamma\lambda}({\bf q},\omega) = \frac{1}{C_{zz}({\bf
q})}\int\limits_{-\infty}^0dt_1 e^{(\epsilon-i\omega) t_1}(I_{s^z}^\gamma({\bf
q});I_{s^z}^\lambda(-{\bf q},t_1)),
 \end{equation}
describes the diffusion of the longitudinal spin respectively along the $D_{zz}^{zz}$  and transverse  $D_{zz}^{xx}$ of the magnetic field, and
 \begin{eqnarray}\label{30}
\nu_{zz}({\bf q},\omega) =  \frac{1}{C_{zz}({\bf q})}\int\limits_{-\infty}^0dt_1
e^{(\epsilon-i\omega)t_1} (\dot{s}^z_{(ms)}({\bf q});\dot{s}^z_{(ms)}(-{\bf q},t_1)),\nonumber\\
 \end{eqnarray}
describes the relaxation of the longitudinal spin components \cite{Bik}. In deriving the last formulas we have taken into consideration the spatial homogeneity of the unperturbed state of the system. At that, the homogeneity exhibits in the vanishing of the correlation functions of the type $(A({\bf q}), B({\bf q}'))$  only if  $({\bf q}'=-{\bf q})$. It is not hard to show that the combination of the correlation functions for $P({\bf q})$  is proportional to the degree of polarization of the electron gas in the normal conductor: $P({\bf q})\sim (n^\up - n^\dw)/ n.$  It should be pointed out that the resulting expressions for the transport coefficients
$D_{zz}^{\gamma\lambda}({\bf q},\omega), \nu_{zz}({\bf q},\omega)$ are suitable for both classical and quantizing magnetic fields and free from assumptions about the nature of the spectrum, the the kind of statistics, etc. Similar expressions for the spin relaxation frequency are also inherent in the theory describing the spin-lattice relaxation of a nonequilibrium spin system with magnetic resonance saturation \cite{Bik}.

 Averaging the operator equation  (\ref{15}),  we have
\begin{eqnarray}\label{31}
<\dot{H}_m({\bf x})> = -\nabla\,<I_{H_m}({\bf x})> + <\dot{H}_{m(ms)}({\bf x})>+\nonumber\\
+<\dot{H}_{m(mp)}({\bf x})>.\qquad\qquad
 \end{eqnarray}
Or averaging time
\begin{eqnarray}\label{33}
 <\overline{\dot{H}_m({\bf x})}> = \int d{\bf x}'\, \,\{ D^{\gamma}_{H_mH_m}({\bf x},{\bf x}')\nabla^\gamma \beta_m({\bf x'}) +
 \nonumber\\+L^m_{m(ms)} ({\bf  x},{\bf x}')\, (\beta_{s}({\bf x}') - \beta_{m}
 ({\bf x}'))+\nonumber\\ +  L^m_{m(mp)}({\bf x},{\bf x}')\,(\beta_{m}({\bf x}') -\beta)\},\qquad\quad
 \end{eqnarray}
where
\begin{eqnarray}\label{34}
L^m_{m(ml)}({\bf x},{\bf x}')  = \int\limits_{-\infty}^0dt'\,e^{\epsilon t'} \,
(\dot{H}_{m(ml)}({\bf x})\,;\,\dot{H}_{m(ml)}({\bf x}',t')),\nonumber\\
 D^{\gamma}_{H_mH_m}({\bf x},{\bf x}')=\int\limits_{-\infty}^0dt'\,e^{\epsilon
t'}\,\nabla^\lambda\,(I^\lambda_{H_m}({\bf x}); I^\gamma_{H_m}({\bf x}',t'))\,.\qquad
 \end{eqnarray}

The first term on the right-hand side of the equation (\ref{31}) is  the energy density change in the magnon subsystem due to the temperature gradient, which in turn leads to the magnon flux. As in the case of the spin electron subsystem, this term describes the magnon diffusion. The second and third terms of the expression (\ref{31}) are responsible for the impact of the electronic (spin) and lattice (phonon) subsystems on the magnon energy change through their interaction. The role of the spin electron subsystem reduces to the generation and annihilation of the magnons by the inelastic spin-flip electron scattering at the normal metal/ferrodielectric interface. According to (\ref{24a}), this contribution is proportional to the spin accumulation $\mu_s$. The phonon subsystem proves to affect the magnon energy change in a twofold manner. On the one hand, the magnon-phonon scattering processes make themselves felt in the energy relaxation behavior of the magnon subsystem, on the other hand, the phonon subsystem often acts as a "heating"\, of the heat/charge transfer processes by means of drag effects \cite{Gur}. It should be emphasized that the booster role can also be studied by the NSO method. The energy transfer efficiency between the phonon and magnon subsystems depends on temperatures of the appropriate subsystems and the degree of their mutual interaction. As can easily be seen from (\ref{33}), a non-zero contribution to the spin-wave current is the consequence of the difference in the temperatures of the above subsystems. The equation (\ref{33}) implies that, depending upon its direction, the magnon flux produced by a temperature gradient can give rise to the angular momentum transfer from the magnon system to the electron system. Besides, the heat fluxes contribute to the emergence both of the thermally induced spin-torque effect \cite{Hat,Jia,Slon} and of the spin-torque effect generated by magnons \cite{Yan,Hin}.

\section*{\bf Conclusions}

Using one of the methods of quantum  nonequilibrium statistical physics (NSO), we investigated the spin transport in hybrid nanostructures: normal metal/ferromagnetic insulator. An approximation of the effective parameters, when each of the interacting subsystems (electron spins, magnons, phonons) is characterized by its effective temperatures was considered. We constructed macroscopic equations for the spin current caused by both the unbalanced spin subsystem and  an inhomogeneous temperature field in the ferromagnetic insulator. We  derived the generalized Bloch equations which describe  the spin and spin-wave current propagation in the system. At that, these allow for both the diffusive nature of the magnon motion and the magnon relaxation processes responsible for the spin pumping
and the spin-torque effect.

\subsection*{appendix}

Let us give the results coming from the expressions for the spin diffusion tensor (\ref{29}), including the temporal and spatial dispersion and find their relationship with the conductivity tensor components. The correlation functions of the fluxes in the formula (\ref{29})  involve the Fourier components of the isothermal Green's functions
\begin{equation}\label{p1}
G_{AB}(t) = \theta(-t)\,e^{\epsilon t}\,(A,\,B(t))_0 =
\int\limits_{-\infty}^{+\infty}\frac{d\omega}{2\pi}\,e^{i\omega t}\,G_{BA}(\omega)
\end{equation}
Differentiating  (\ref{p1}) over $ t$, we obtain the chain of equations
\begin{eqnarray}\label{p2}
 \left(\frac{\partial}{\partial t}-\epsilon -i\omega_0\right)\,G = - \delta(t)(B,A) + G_1,\nonumber\\
 \left(\frac{\partial}{\partial t}-\epsilon -i\omega_0\right)\,G_1 = -\delta(t)(B,\dot{A}) - G_2,\nonumber\\
 \ldots\ldots\ldots \quad,
\end{eqnarray}
$G(t)=G_{AB}(t),\quad  G_1(t) = \theta(-t)\,e^{\epsilon t}\,(A,\,\dot{B}(t))_0.$
$$ G_2(t) = \theta(-t)\,e^{\epsilon t}\,(\dot{A},\,\dot{B}(t))_0.$$
Here  $\dot{A}=(i\hbar)^{-1}[A,H_V],\quad H_V$ is the scatterer Hamiltonian. The formal solution of the chain is
\begin{equation}\label{p3}
 G_{AB}(\omega) = [M_{AB}(\omega) + \epsilon - i(\omega-\omega_0)]^{-1}(A,B),
\end{equation}
where $M(\omega)=G_1\,G^{-1}$ is the mass operator for the Green's function.

Let us consider the frequency dispersion by the example of the longitudinal spin diffusion coefficient. Restricting ourselves to the Born approximation of the interaction with the lattice in the expression for the mass operator, we have
\begin{equation}\label{p4}
D_{zz}^{zz} = \frac{1}{C_{zz}}\,\frac{(I_{s^z}^z,I^z_{s^z})_0}{\nu_{zz}^{zz}(\omega) - i\omega}.
 \end{equation}
Calculating the correlation functions $(A,B)_0$, we come up with
$$(I_{s^z}^z,I^z_{s^z})_0 = n / 2m ,\quad  \quad C_{zz} =\frac{n}{8}\frac{F_{-1/2}(\mu/T)}{F_{1/2}(\mu/T)},$$
where  $F_m(x)$ are the Fermi integrals. The expression  $ \nu_{zz}^{zz}$  coincides exactly with the formula for the relaxation frequency $\nu_p$  of the electron momentum  \cite{Kal-1}. Thus, the longitudinal spin diffusion coefficient is given by
\begin{equation}\label{p4}
D_{zz}^{zz} = D_0\,\frac{\nu_p}{\nu_p - i\omega},\qquad D_0 =
\frac{2T}{\nu_p}\frac{F_{-1/2}(\mu/T)}{F_{1/2}(\mu/T)}.
 \end{equation}
The components of the spin diffusion tensor can be expressed in terms of the components of the conductivity tensor $\sigma_{ik}({\bf q},\omega)$, which in our notation is
\begin{equation}\label{p5}
\sigma_{\gamma k} = \frac{e^2}{T}\int\limits_{-\infty}^0 dt e^{(\epsilon-i\omega)t}\,(I_N^\gamma(q),
I_N^k(-q,t)).
 \end{equation}
Simple calculations show that
$$D^{zz}_{ik}({\bf q},\omega) = \frac{T}{4e^2\,C_{zz}(q)}\,\sigma_{ik}({\bf q},\omega).$$

The given work has been done as the part of the state task on the theme "Spin" 01201463330 (project 12-T-2-1011) with the support of the Ministry of Education of the Russian Federation (Grant 14.Z50.31.0025)

\end{document}